\documentclass[aps,pre,twocolumn,showpacs,%
preprintnumbers,amssymb]{revtex4-1}
\usepackage{graphicx}
\usepackage[colorlinks=true,linkcolor=blue,%
pagecolor=blue,citecolor=blue,%
urlcolor=blue,anchorcolor=black,%
bookmarksnumbered=true,%
bookmarksopen=true]{hyperref}
\usepackage{dcolumn}
\usepackage{bm}
\renewcommand{\frac}[2]{\displaystyle{#1 \over #2}}
\begin{document}
\title{Dust acoustic waves in three-dimensional 
complex plasmas with a similarity property }
\author{D.~I.~Zhukhovitskii} \email{dmr@ihed.ras.ru}
\homepage{http://oivtran.ru/dmr/}
\affiliation{Joint Institute of High Temperatures, Russian 
Academy of Sciences, Izhorskaya 13, Bd.~2, 125412 
Moscow, Russia}
\date{\today}
\begin{abstract}
Dust acoustic waves in the bulk of a dust cloud in complex 
plasma of low-pressure gas discharge under microgravity 
conditions are considered. The complex plasma is assumed to 
conform to the ionization equation of state (IEOS) developed 
in our previous study. This equation implies the ionization 
similarity of plasmas. We find singular points of IEOS that 
determine the behavior of the sound velocity in different 
regions of the cloud. The fluid approach is utilized to deduce 
the wave equation that includes the neutral drag term. It is 
shown that the sound velocity is fully defined by the particle 
compressibility, which is calculated on the basis of the used 
IEOS. The sound velocities and damping rates calculated for 
different three-dimensional complex plasmas both in ac and 
dc discharges demonstrate a good correlation with 
experimental data that are within the limits of validity of the 
theory. The theory provides interpretation for the observed 
independence of the sound velocity on the coordinate and for 
a weak dependence on the particle diameter and gas pressure. 
Predictive estimates are made for the ongoing PK-4 
experiment.
\end{abstract}
\maketitle
\section{\label{s1}INTRODUCTION}

A low-temperature plasma, which includes dust particles 
with sizes ranging from $1$
 to $10^3 \;\mu {\mbox{m}}$, is usually referred to as 
dusty or complex plasma. Since the mobility of electrons is 
much greater than that of ions, particles acquire a 
significant negative electric charge. This leads to formation 
of a strongly coupled plasma \cite{1,2,3,4,5,6,8,9,47}, in 
which various collective phenomena at the level of 
individual particles can be observed. Complex plasmas are 
studied in gas discharges at low pressures, e.g., in radio 
frequency (RF) discharges. Under microgravity conditions, 
large volumes of three-dimensional (3D) complex plasma 
can be observed. These conditions are realized either in 
parabolic flights \cite{10,11,12,13,14} or onboard the 
International Space Station (ISS) 
\cite{10,15,16,17,18,019,19}.

The PK-4 project is intended to be a continuation of 
successful series of PK-1, PK-2, PK-3, and PK-3 Plus 
experiments onboard the ISS. The PK-4 setup was 
exhaustively tested in ground-based conditions \cite{60} 
and in parabolic flights \cite{37,38,39}. Since the PK-4 
experiments are focused on dynamical phenomena in 
complex plasmas including formation and propagation of 
the shock waves and solitons, investigation of the waves 
associated with the motion of dust particles is of special 
interest. The linear waves with a long wavelength larger 
than the interparticle distance and the Debye length are 
commonly called the dust acoustic waves (DAWs).

From the time that the notion of DAWs was first introduced 
by Rao, Shukla, and Yu \cite{48}, DAWs became a subject 
of extensive investigations \cite{36}. Havnes {\it et al.\/} 
\cite{25} predicted that if the disturbance generating DAWs 
in a strongly coupled system of the dust particles moves 
with a supersonic velocity then the Mach cone emerges. 
The Mach cone observations were used for the 
determination of dust sound velocity, first, in the 
experiments with a 2D lattice plane 
\cite{26,21,30,27,28,31,32} and then in a 3D strongly 
coupled system of charged particles \cite{11,12,019,19,14} 
formed in RF discharge in argon. In Ref.~\cite{35}, the 
Mach cone observation in a complex plasma of the neon RF 
discharge was used for the determination of the dust sound 
velocity. An interesting feature of the 3D studies was 
independence of the sound velocity (within the 
experimental error) of the system parameters such as the 
dust particle radius, argon pressure, and the location in the 
bulk of a dust cloud. The latter fact is most surprising 
because according to the assessment \cite{22}, the particle 
number density in the inner and outer regions of the dust 
cloud differs almost by an order of magnitude; the particle 
charge also changes significantly along the bulk of the 
cloud. In addition, the sound velocity was found to be 
isotropic, i.e., it is independent of the direction of wave 
propagation with respect to the direction of the gas 
discharge electric field. In experiments with argon 
\cite{11,12,019,19,14}, the measured sound velocity ranged 
from $2$
 to $3\;{\mbox{cm/s}}$; the sound velocity measured in 
neon proved to be twice as low (about 
$1\;{\mbox{cm/s}}$) \cite{35}, but it was still independent 
of the particle number density.

In the pioneer work \cite{48}, calculation of the dust sound 
velocity was based on the fluid approach. This result is 
valid if the dust component of the complex plasma 
conforms to the ideal gas equation of state while the 
average kinetic energy of particles (particle temperature 
$T_d $) is equal to zero. The resulting formula for the 
sound velocity is similar to that for the ion acoustic waves 
\cite{24}. However, the systems of dust particles, for which 
the sound velocity was measured, are strongly coupled with 
a typical coupling parameter $\Gamma = Z^2 e^2 /r_d T_d 
> 200$
 \cite{49}, where $Z$
 is the dust particle charge in units of the electron charge, 
$e$
 is the elementary electric charge, $r_d = (3/4\pi n_d 
)^{1/3}$ is the Wigner--Seitz radius for the dust particles, 
$n_d$ is the particle number density, and the Boltzmann 
constant is set to unity. Note that this estimate is most likely 
a low bound for $\Gamma $, which relates to the smallest 
particles used in experiments and to the lowest estimates for 
the particle charge; for real systems, $\Gamma$ can be 
orders of magnitude higher. Note that the dust kinetic 
temperature determined experimentally is rather high: 
according to Ref.~\cite{50}, $T_d < 0.8\;{\mbox{eV}}$; in 
Ref.~\cite{51}, the dust temperature was found to be in the 
range from $0.1$
 to $1\;{\mbox{eV}}$. Thus, $T_d /T_n \sim 30$, where 
$T_n$ is the temperature of gas molecules usually assumed 
to be equal to room temperature. In spite of this fact, 
$\Gamma \gg 1$.

Khrapak and Thomas \cite{40} showed that for strongly 
coupled Yukawa systems, the dust sound propagation is 
defined by the nonzero compressibility of the particles 
resulting from their correlation energy. The term in the 
expression for the sound velocity corresponding to the 
electric field perturbation due to charge separation in the 
sound wave, which was solely taken into account in 
Ref.~\cite{48}, appeared to be canceled out exactly by the 
plasma-related contribution to the isothermal 
compressibility modulus. However, the results of this work 
are not directly applicable to the complex plasmas because 
the latter is different from the Yukawa systems 
\cite{45,46}. A principal difference comes from the fact 
that the complex plasma is an open system characterized by 
an intense energy exchange with the environment; due to 
the production of electrons and ions, their recombination 
and loss on chamber walls, and charging of the dust 
particles, the charged components of complex plasma do 
not form a fixed ensemble. Therefore, equilibrium 
thermodynamics along with all related thermodynamic 
notions such as the entropy, {\it is not applicable\/} for this 
system. Another difference lies in the presence of an 
external electric field typical for both the ac and dc 
discharges. This field is responsible for the emergence of 
ion drag force, which along with the electric field defines 
the dynamics of dust particles \cite{22}. It is also important 
that the particle charge is not fixed but is a function of the 
local number densities of electrons, ions, and particles, 
whose variation may reach an order of magnitude. Note that 
the particle charge variation was not allowed for in 
Refs.~\cite{48} and \cite{40}. Nevertheless, the real 
complex plasma can be {\it locally\/} treated as a system of 
strongly correlated particles on a uniform charge 
background formed by the electrons and ions, i.e., as the 
one-component plasma. This determines a principal 
similarity between the complex plasma and the strongly 
coupled Yukawa system.

The objective of this work is to develop a theory of DAWs 
propagation in a real system (the dust cloud in a 
low-pressure discharge). Our approach is based on the 
model of complex plasma \cite{22}. According to this 
model, the dust cloud can be stable if the electric force from 
an external electric field is balanced by the force due to 
collisions with the ions drifting in this electric field (ion 
drag force). The key assumption of the model is 
overlapping of the Coulomb potentials of neighboring 
particles, due to which the cross section of ion scattering on 
the particles is a function of the particle number density. 
This made it possible to relate the number density of 
electrons, ions, and particles, and the particle charge in such 
a way that each parameter defines uniquely the others. The 
``dust invariant'' was obtained that is nearly constant for 
the dust clouds observed in different experiments. As a 
consequence, a reasonable estimation for the stationary 
particle number density, which was shown to be 
independent of the number of injected particles, was 
obtained. The resulting ionization equation of state (IEOS) 
was written in two dimensionless variables. As follows 
from this IEOS, for two different systems, the ratio of 
characteristic quantities in corresponding ionization states 
is the same as that at the critical points (generally, at all 
singular points). For this reason, one can call such plasmas 
similar.

The propagation of DAWs is investigated using the fluid 
approach. Note that the Navier--Stokes equation can be 
applied for the collective motion of dust particles even at 
the length scales commensurable with several interparticle 
distances \cite{49,52,53}. The same approach was used for 
the strongly coupled Yukawa systems in Ref.~\cite{40}. In 
the fluid dynamics equations, we take into account the 
electric force and the ion and the neutral drag. Linearization 
of these equations shows that the balance of forces in an 
unperturbed medium leads to full compensation of the 
contribution from the electric field perturbation. Thus, the 
sound velocity is solely defined by the dust compressibility, 
which can be calculated on the basis of used IEOS. Here, 
the situation is similar to that for the strongly coupled 
Yukawa systems \cite{40}. In addition, we allow for the 
interaction between the particles and the neutrals (neutral 
drag), which makes it possible to calculate the damping rate 
of DAWs.

The resulting formula for the dust sound looks quite 
different from that of Ref.~\cite{48} (and also, from 
Ref.~\cite{40}). In accordance with IEOS, an increase of 
the particle number density entails a decrease in the particle 
charge. Consequently, the dust compressibility and the 
sound velocity prove to be almost constant everywhere in 
the bulk of dust cloud, in spite of a considerable variation 
of the complex plasma parameters. The sound velocity 
proves to depend weakly on the dust particle radius and gas 
pressure. For the experiments with argon, it is in a 
reasonable agreement with the experimental data, in 
contrast to the calculation using the formula obtained in 
Ref.~\cite{48}. Developed theory makes it possible to 
perform predictive estimations for conditions typical for the 
PK-4 experiments.

The paper is organized as follows. In Sec.~\ref{s2}, we 
formulate the governing equations for similar complex 
plasmas. In Sec.~\ref{s3}, we represent IEOS for the 
stationary dust cloud in the one-parametric form and 
explore its singular points, which are the boundaries of 
characteristic behavior regions of the sound velocity. In 
Sec.~\ref{s4}, we obtain the DAWs' dispersion relation and 
calculate the sound velocity by derivation of the dust 
compressibility. We compare the calculation results with 
available experimental data in Sec.~\ref{s5}. The results of 
this study are summarized in Sec.~\ref{s6}.

\section{\label{s2}THE FLUID APPROACH TO 
COMPLEX PLASMA}

Consider a dust cloud in the low-pressure gas discharge. 
Under microgravity conditions (either in parabolic flights or 
onboard the ISS), a dust particle is subject to three basic 
forces: the electric driving force, the ion drag force arising 
from scattering of the streaming ions on dust particles, and 
the neutral drag force (friction force) due to collisions of 
the atoms with the moving particles. Note that in a strongly 
coupled system, the correlation energy originating from 
particle ordering results in the difference between the 
volume-averaged electric field and the electric field at the 
point of particle location. This effect can be included if we 
introduce the dust pressure. The effect of this pressure on 
the force balance equation in a stationary plasma is 
discussed in Sec.~\ref{s3}, where we consider the 
stationary (unperturbed) state of the dust cloud.

If we represent the dust component of complex plasma as a 
fluid, we have the following basic equations. The first is the 
Euler equation
\begin{equation}
\frac{{\partial {\bf{v}}}}{{\partial t}} + 
({\bf{v}}\cdot\nabla ){\bf{v}} + \nu {\bf{v}} = 
\frac{1}{\rho }({\bf{f}}_e + {\bf{f}}_{id} - \nabla p), 
\label{e1}
\end{equation}
where ${\bf{v}}(t,\,{\bf{r}})$
 is the velocity field, $p$
 and $\rho$ are the pressure and density of a fluid, 
respectively; $\rho \nu {\bf{v}}$
 is the neutral drag force acting on unit volume, $\nu = 
(8\sqrt {2\pi } /3)\delta m_n n_n v_{T_n } a^2 /M$
 is the friction coefficient \cite{33,6}, $\delta \simeq 1.4$
 is the accommodation coefficient; $m_n$ is the mass of a 
gas molecule; $n_n$ and $v_{T_n } = (T_n /m_n )^{1/2}$ 
are the number density and thermal velocity of the gas 
molecules, respectively, $T_n = 300\;{\mbox{K}}$
 is the temperature of a gas; $a$
 is the particle radius, $M = (4\pi /3)\rho _d a^3$ is its 
mass; $\rho _d$ is the particle material density; and 
${\bf{f}}_e$ and ${\bf{f}}_{id}$ are the electric field 
driving force and the ion drag force acting on unit volume, 
respectively. Here,
\begin{equation}
{\bf{f}}_e = - Zen_d {\bf{E}} = - \frac{{aT_e }}{e}\Phi 
n_d {\bf{E}}, \label{e4}
\end{equation}
where $T_e$ is the electron temperature, $\Phi = - Ze^2 
/aT_e$ is the dimensionless potential of a dust particle, 
${\bf{E}} = (T_e /e)\nabla \ln n_e$ is the electric field 
strength, $n_e$ is the electron number density, and
\begin{equation}
{\bf{f}}_{id} = \frac{3}{8}\left( {\frac{{4\pi }}{3}} 
\right)^{1/3} n_d^{1/3} n_i \lambda e{\bf{E}}, \label{e5}
\end{equation}
where $\lambda$ is the ion mean free path with respect to 
the collisions against gas atoms, and $n_i$ is the electron 
number density. Equation (\ref{e5}) is based on a simple 
estimation of the momentum transfer from the ions to the 
particle for the case of overlapping Coulomb potentials of 
neighboring particles \cite{22}. Note that this equation is 
invalid for an isolated particle.

In addition, the fluid of dust particle obeys the continuity 
equation
\begin{equation}
\frac{{\partial \rho }}{{\partial t}} + \nabla \cdot (\rho 
{\bf{v}}) = 0. \label{e2}
\end{equation}
For the dust pressure, we use the expression derived in 
Ref.~\cite{22}
\begin{equation}
p = \frac{1}{{8\pi }}\left( {\frac{{aT_e }}{{e\lambda ^2 
}}} \right)^2 p^* ,\;\;p^* = \Phi ^2 n^{*4/3} , \label{e3}
\end{equation}
where $n^* = (4\pi /3)\lambda ^3 n_d$ and $n_d = \rho 
/M$
 is the particle number density (we mark dimensionless 
quantities with an asterisk). The expression (\ref{e3}) can 
be rewritten in the form $p = Z_{\mathrm{th}} n_d T_d $, 
where $Z_{\mathrm{th}} = \Gamma /6$
 is the compressibility factor for the dust. Since $\Gamma 
\gg 1$
 even for $T_d \gg T_n $, we have $Z_{\mathrm{th}} \gg 
1$; practically, $Z_{\mathrm{th}} > 30$. This agrees 
qualitatively with the results of calculation for the strongly 
coupled Yukawa system \cite{44}. Hence, the thermal 
equation of state for the dust component is fully defined by 
its strong coupling. Equation~(\ref{e3}) results in a typical 
dust pressure of $10^{ - 7} \div 10^{ - 6} \;{\mbox{Pa}}$, 
which is in a good agreement with its estimate following 
from the assessment of the deformation threshold for a 
cavity around a large particle in a complex plasma \cite{53} 
and of the radius of such cavity \cite{22}. The same orders 
of magnitude of the dust pressure were determined in 
Ref.~\cite{50}, where the dust acoustic shock waves in 
strongly coupled dusty plasma were investigated 
experimentally.

\section{\label{s3}SINGULAR POINTS OF IEOS FOR 
A STATIONARY DUST CLOUD}

As follows from Eq.~(\ref{e1}), the stationary state 
condition for a dust cloud (${\bf{v}} \equiv 0$) is
\begin{equation}
{\bf{f}}_e + {\bf{f}}_{id} - \nabla p = 0. \label{e6}
\end{equation}
It can be easily shown that the third term on the l.h.s.\ of 
Eq.~(\ref{e6}) is much smaller than ${\bf{f}}_e$ and 
${\bf{f}}_{id} $. Indeed, this term can be represented in 
the form
\begin{equation}
\nabla p = \frac{{dp}}{{d\rho }}\nabla \rho = c^2 \nabla 
\rho = Mn_d c^2 \nabla \ln n_d , \label{e7}
\end{equation}
where $c^2 = dp/d\rho $. It is shown below that $c$
 is defined by ordinary rather than partial derivative and its 
physical meaning as the sound velocity is clarified in 
Sec.~\ref{s4}. Since ${\bf{f}}_e = - Zn_d T_e \nabla \ln 
n_e $, where $\left| {\nabla \ln n_e } \right| \sim 1/L$
 and $L$
 is the length of the dust cloud, we have $\left| {{\bf{f}}_e 
} \right| \sim an_d \Phi T_e^2 /Le^2$ not too close to the 
void boundary. Hence, we obtain from Eq.~(\ref{e7}) and 
the estimation $\left| {\nabla \ln n_d } \right| \sim 1/L$
 that the condition $\left| {{\bf{f}}_e } \right| \gg \left| 
{\nabla p} \right|$
 is satisfied if
\begin{equation}
\frac{{a\Phi T_e^2 }}{{Mc^2 e^2 }} \gg 1. \label{e8}
\end{equation}
For the conditions of experiment \cite{19}, the l.h.s.\ of 
(\ref{e8}) is of the order of $10^3$ (if $c = 
2.4\;{\mbox{cm/s}}$).

Thus, Eq.~(\ref{e6}) is reduced to ${\bf{f}}_e + 
{\bf{f}}_{id} = 0$
 or
\begin{equation}
\frac{\pi }{2}r_d^2 n_i \lambda = \frac{{aT_e }}{{e^2 
}}\Phi , \label{e009}
\end{equation}
which coincides with the balance equation (1) of 
Ref.~\cite{22}. The combination of this equation with the 
equation for particle potential that follows from the orbital 
motion limited (OML) approximation \cite{54,55} at $T_e 
/T_i \gg 1$
\begin{equation}
\theta \Phi e^\Phi  = \frac{{n_e }}{{n_i }}, \label{e9}
\end{equation}
where $\theta = \sqrt {T_e m_e /T_i m_i } $, $T_i \approx 
T_n$ and $m_i$ are the ion temperature and mass, 
respectively, and $m_e$ is the electron mass, and the local 
quasineutrality condition
\begin{equation}
n_i = \frac{{aT_e }}{{e^2 }}\Phi n_d + n_e , \label{e10}
\end{equation}
yields the stationary IEOS
\begin{equation}
\theta \Phi e^\Phi  + \frac{3}{8}\left( {\frac{{\pi n_i^* 
}}{{2\Phi }}} \right)^{1/2} = 1, \label{e11}
\end{equation}
where $n_i^* = (e^2 \lambda ^3 /aT_e )n_i $. In contrast to 
IEOS (1) of Ref.~\cite{22}, (\ref{e11}) includes a single 
parameter $\theta $. The solution of 
Eqs.~(\ref{e009})--(\ref{e10}) can be represented in the 
form of a one-parametric relation between each desired pair 
of the variables $n_i $, $n_e $, $n_d $, and $\Phi$ (each 
relation is IEOS as well), e.g., the relation between $n_d$ 
and $\Phi $. Note that temperatures are included in the 
parameter $\theta$ because they are assumed to be fixed. 
Thus, for treated system, the dust compressibility is 
proportional to ordinary rather partial derivative $dp/d\rho 
$, which implies that differentiation is performed along the 
ionization equilibrium line.
\begin{figure}
\centering \unitlength=0.24pt
\begin{picture}(800,800)
\put(-130,-20){\includegraphics[width=9.7cm]{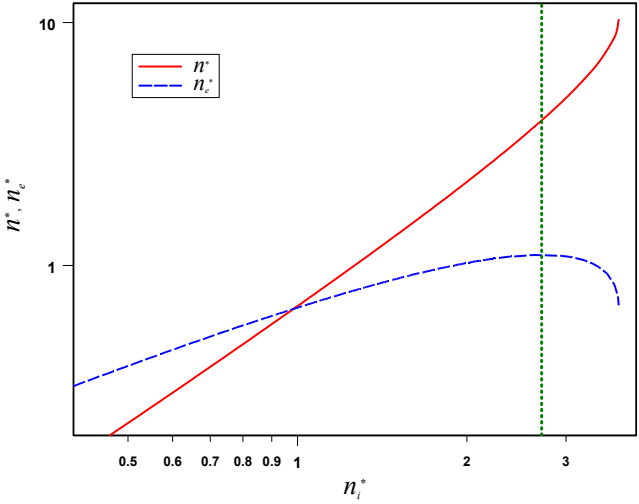}}
\end{picture}
\caption{\label{f1} (Color online) Dimensionless dust 
particle ($n^* $, solid line) and electron number density 
($n_e^* $, dashed line) as a function of the dimensionless 
ion number density $n_i^* $, $\theta = 
{\mbox{0}}{\mbox{.0431}}$.}
\end{figure}

Figure~\ref{f1} illustrates two IEOS' in the variables 
$n_i^* $, $n^*$ and $n_i^* $, $n_e^* $, where $n_e^* = 
(e^2 \lambda ^3 /aT_e )n_e$ is the dimensionless electron 
number density, for typical conditions of the PK-3 Plus 
experiment \cite{19}. The dependence $n_e^* (n_i^* )$
 shows that both $n_i^*$ and $n_e^*$ have a maximum, 
which corresponds to two singular points of IEOS. The 
maximum ion number density $n_{ic}^*$ is reached at the 
first singular point. Here, $n^*$ has a maximum 
(Fig.~\ref{f1}). The first singular point, which is similar to 
the critical point and can be associated with the void 
boundary, is defined by the condition $\left. {(dn_i^* 
/dn_e^* )} \right|_{n_e^* = n_{ec}^* } = 0$. Since $\left. 
{(d\Phi /dn_e^* )} \right|_{n_e^* = n_{ec}^* } \ne 0$, this 
condition can be rewritten as $\left. {(dn_i^* /d\Phi )} 
\right|_{\Phi = \Phi _c } = 0$
 (see Fig.~4 in Ref.~\cite{22}). We substitute $n_i^*$ from 
(\ref{e11}) in the latter derivative to obtain the equation 
defining the potential $\Phi _c$ at the first singular point
\begin{equation}
\theta \Phi _c e^{\Phi _c } (2\Phi _c + 3) = 1. \label{e12}
\end{equation}
Its solution is
\begin{equation}
\Phi _c \simeq - \ln \theta - \ln \frac{{\Phi _0 }}{2} - \ln 
(\Phi _0 + 3), \label{e13}
\end{equation}
where $\Phi _0$ is the potential of an isolated particle 
obtained from (\ref{e9}) with $n_e = n_i $: $\theta \Phi _0 
e^{\Phi _0 } = 1$. An approximate solution of this equation 
is $\Phi _0 \simeq - \ln \theta - \ln ( - \ln \theta - 1)$. From 
(\ref{e11}) and (\ref{e12}) we obtain the ion number 
density at the first singular point (critical number density)
\begin{equation}
n_{ic}^* = \frac{{512}}{{9\pi }}\Phi _c \left( {\frac{{\Phi 
_c + 1}}{{2\Phi _c + 3}}} \right)^2 . \label{e14}
\end{equation}

Given $n_{ic}^* $, one can calculate from 
Eqs.~(\ref{e9})--(\ref{e11}) the dimensionless electron 
number density $n_{ec}^*$ at the first singular point. With 
due regard for (\ref{e12}), we obtain
\begin{equation}
n_{ec}^* = \frac{{512}}{{9\pi }}\Phi _c \frac{{(\Phi _c + 
1)^2 }}{{(2\Phi _c + 3)^3 }}. \label{e15}
\end{equation}
In a similar way, we obtain from (\ref{e10}) the particle 
number density at the first singular point
\begin{equation}
n_c^* = \frac{{4096}}{{27}}\left( {\frac{{\Phi _c + 
1}}{{2\Phi _c + 3}}} \right)^3 . \label{e16}
\end{equation}
Note that $n_c^*$ is the maximum of $n^*$ at fixed 
$\theta $. At $\theta \to \infty$ and $\Phi _c \to \infty $, 
$n_c^* \to 512/27$, which is the upper bound for this 
quantity. The Havnes parameter $H = Zn_d /n_e = (3/4\pi 
)(\Phi n/n_e^* )$
 is introduced to quantify the fraction of negative charge 
accumulated on the particles. At the first singular point, 
$H_c = 2(\Phi _c + 1)$.

\begin{figure}
\centering \unitlength=0.24pt
\begin{picture}(800,800)
\put(-145,-20){\includegraphics[width=9.85cm]{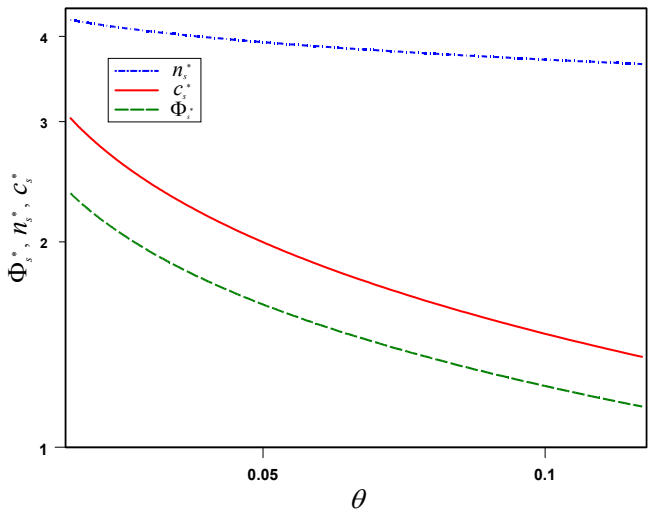}}
\end{picture}
\caption{\label{f2} (Color online) Dimensionless quantities 
at the second singular point as a function of the temperature 
parameter $\theta $. Solid line indicates the sound velocity 
$c_s^* $; dashed line, the dust particle potential $\Phi _s^* 
$; and dashed-dotted line, the particle number density 
$n_s^* $.}
\end{figure}

We exclude $n_i$ from (\ref{e9}) and (\ref{e11}) to derive 
IEOS in the variables $n_e$ and $\Phi$\begin{equation}
n_e^* = \frac{{128}}{{9\pi }}\theta \Phi ^2 e^\Phi (1 - 
\theta \Phi e^\Phi )^2 . \label{e17}
\end{equation}
The second singular point corresponds to the maximum of 
$n_e$ (Fig.~\ref{f1}). We denote the quantities 
corresponding to this point by subscript $s$. The second 
singular point is defined by the condition $\left. {(dn_e^* 
/dn_i^* )} \right|_{n_i^* = n_{is}^* } = 0$. Since $\left. 
{(dn_i^* /d\Phi )} \right|_{\Phi = \Phi _s } \ne 0$, we 
rewrite it in the form $\left. {(dn_e^* /d\Phi )} \right|_{\Phi 
= \Phi _s } = 0$. This yields the equation for the potential 
$\Phi _s$\begin{equation}
\frac{{2\theta e^{\Phi _s } (\Phi _s + 1)}}{{1 - \theta \Phi 
_s e^{\Phi _s } }} = 1 + \frac{2}{{\Phi _s }}. \label{e18}
\end{equation}
It is seen from Eq.~(\ref{e18}) that $\Phi _s \simeq - \ln 
3\theta$ at $\theta \to 0$
 and $\Phi _s \simeq 1/2\theta$ at $\theta \to \infty $. The 
decreasing behavior of $\Phi _s (\theta )$
 is illustrated by Fig.~\ref{f2}, in which the solutions of 
Eq.~(\ref{e18}) for different $\theta$ are shown.

The combination of (\ref{e17}) and (\ref{e18}) yields the 
electron number density $n_{es}^*$ at the second singular 
point
\begin{equation}
n_{es}^* = \frac{{512}}{{9\pi }}\Phi _s \frac{{(\Phi _s + 
2)(\Phi _s + 1)^2 }}{{(3\Phi _s + 4)^3 }}. \label{e19}
\end{equation}

One can estimate the ion number density $n_{is}^*$ using 
(\ref{e9}) and (\ref{e19}),
\begin{equation}
n_{is}^* = \frac{{512}}{{9\pi }}\Phi _s \left( {\frac{{\Phi 
_s + 1}}{{3\Phi _s + 4}}} \right)^2 . \label{e20}
\end{equation}

We substitute $n_e /n_i$ from (\ref{e9}) into (\ref{e10}) 
and then $n_i$ from (\ref{e10}) into (\ref{e11}) to derive 
IEOS in the variables $n^*$ and $\Phi $:
\begin{equation}
n^* = \frac{{512}}{{27}}(1 - \theta \Phi e^\Phi )^3 . 
\label{e21}
\end{equation}
It follows from (\ref{e18}) and (\ref{e21}) that
\begin{equation}
n_s^* = \frac{{4096}}{{27}}\left( {\frac{{\Phi _s + 
1}}{{3\Phi _s + 4}}} \right)^3 . \label{e22}
\end{equation}
It is seen in Fig.~\ref{f2} that the particle number density at 
the second singular point $n_s^* \approx 4$, and it is 
almost independent of $\theta $. At this point, the Havnes 
parameter is
\begin{equation}
H_s = 2\frac{{\Phi _s + 1}}{{\Phi _s + 2}}. \label{e023}
\end{equation}
For example under conditions of the experiment \cite{19}, 
$\theta = {\mbox{0}}{\mbox{.0431}}$, $\Phi _s = 1.708$, 
and we obtain from (\ref{e023}) $H_s = 1.461$.

Strictly speaking, at the second singular point, the balance 
equation (\ref{e009}) is invalid along with IEOS 
(\ref{e11}) and (\ref{e17}) because $\nabla n_e = 0$
 and the electric field ${\bf{E}}$
 vanishes. However, as follows from (\ref{e8}), the width 
of singular region is of the same order of magnitude as the 
interparticle distance. At this point, ${\bf{E}}$
 must change its direction. It is possible that this region 
corresponds to that of inverse ion streaming considered in 
Ref.~\cite{56}.

In this Section, we have generalized and extended the 
results of the previous study \cite{22}. We have shown that 
IEOS for the dust cloud has two singular points, which 
define the maximum ion and electron number density, 
respectively, that can be attained in a spatial region 
occupied by the cloud. The complex plasma parameters at 
these points are typical for performed experiments. For 
each singular point, we have obtained analytical 
expressions for the dimensionless number densities of the 
electrons, ions, and particles, as well as for the particle 
electrostatic potential and the Havnes number. We have 
demonstrated that all these quantities are functions of a 
single quantity, the dimensionless particle potential $\Phi$ 
at a corresponding point. The latter is defined by the 
parameter $\theta $, which is a function of the gas 
molecular mass and the electron temperature.

The spatial location of two singular points can be illustrated 
in application to the dust cloud in RF discharge plasma 
(e.g., PK-3 Plus \cite{18} or IMPF-K2 \cite{11} 
experiments). The configuration of a dust cloud can be very 
crudely regarded as a sphere with an empty spherical region 
in the center (void). Then we can associate the first singular 
point with the void boundary and the second singular point, 
with a sphere of the radius larger than that of the void. 
Thus, we have two spatial regions separated by spherical 
surfaces: the region between the first and the second 
singular points and the region outside the second singular 
point.

\section{\label{s4}DUST ACOUSTIC WAVES AND 
THE SOUND VELOCITY}

In what follows, we will treat a nonstationary solution of 
Eqs.~(\ref{e1}) and (\ref{e2}) corresponding to DAWs. As 
usual, we imply that the wavelength $2\pi c/\omega $, 
where $\omega$ is the frequency, is the largest length scale 
of the problem and that the wave propagates adiabatically 
and can be treated in the linear approxiamtion. First, we 
represent the sum ${\bf{f}}_e + {\bf{f}}_{id}$ in 
(\ref{e1}) in the form $(g_e + g_{id} ){\bf{E}}$, where 
$g_e = - aT_e \Phi n_d /e$
 and $g_{id} = (3/8)(4\pi /3)^{1/3} n_d^{1/3} n_i \lambda 
e$
 [cf.\ (\ref{e4}) and (\ref{e5})]. Then we write $g_e = 
g_{e0} + g'_e $, $g_{id} = g_{id0} + g'_{id} $, and 
${\bf{E}} = {\bf{E}}_0 + {\bf{E'}}$, where $g_{e0} $, 
$g_{id0} $, and ${\bf{E}}_0$ are the unperturbed 
quantities and $g'_e $, $g'_{id} $, and ${\bf{E'}}$, the 
perturbed ones. Assuming a small perturbation, we write
\begin{equation}
\begin{array}{*{20}c}
  {{\bf{f}}_e + {\bf{f}}_{id} \simeq (g_{e0} + g_{id0} 
){\bf{E}}_0 + (g_{e0} + g_{id0} ){\bf{E'}}} \\
  {\quad \;\;\,\, + (g'_e + g'_{id} ){\bf{E}}_0 = (g'_e + 
g'_{id} ){\bf{E}}_0 } \\
\end{array} \label{e099}
\end{equation}
due to the force balance condition $g_{e0} + g_{id0} = 0$. 
It is noteworthy that this condition {\it cancels out 
exactly\/} the field perturbation ${\bf{E'}}$. In other terms, 
although the field perturbation is nonzero, it does not 
contribute to the equation of fluid dynamics (\ref{e1}). 
Therefore, the effect of charge separation due to the ion 
shift that determines the sound velocity in equilibrium 
plasma \cite{48} is fully compensated in the complex 
plasma of gas discharge treated in this work. Thus, the 
sound velocity is solely defined by the dust compressibility. 
A similar situation takes place in strongly coupled Yukawa 
systems \cite{40}.

Calculation of the sum $g'_e + g'_{id}$ in (\ref{e099}) is a 
separate problem, which cannot be solved within the 
framework of the model \cite{22} we use in this work. In 
fact, the model assumes that at least one stationary spatial 
distribution of the number densities of the charged 
components (electrons, ions, and particles) is known. 
Hence, to evaluate $g'_e + g'_{id} $, we have either to 
develop a self-consistent theory of the complex discharge 
plasma (which is now lacking) or to add another 
assumption to our model, in other words, to redefine the 
model so that it would be applicable for dynamic processes. 
Note that the models of equilibrium plasma and of strongly 
coupled Yukawa systems are also defined by a set of 
assumptions. Our additional assumption will be $g'_e = - 
g'_{id} $, which means simply that the variations of all 
quantities in the sound wave are related by IEOS 
(\ref{e11}) or that the electric and ion drag forces are 
always kept in balance. This assumption is similar to that 
made for an ordinary acoustic wave. What is more 
important, this assumption allows one to provide an 
interpretation of the sound velocity isotropy known from 
experiment. Indeed, the sound velocity was found to be 
independent of the wave propagation direction with respect 
to the direction of external electric field ${\bf{E}}_0 $, 
which is almost radial in some setups. If only $g'_e \ne - 
g'_{id} $, the resulting sound velocity would be 
anisotropic.

With this assumption, Eq.~(\ref{e1}) is reduced to
\begin{equation}
\frac{{\partial {\bf{v}}}}{{\partial t}} + 
({\bf{v}}\cdot\nabla ){\bf{v}} + \nu {\bf{v}} = - 
\frac{1}{\rho }\nabla p. \label{e23}
\end{equation}
Actually, (\ref{e23}) corresponds to a fluid of soft spheres 
with no Coulomb interaction. This equation was 
successfully applied for the problem of dust particle 
collective dynamics \cite{49,52,53}. Equation (\ref{e23}) 
differs from the standard wave equation by the neutral drag 
force term on its l.h.s. We linearize Eqs.~(\ref{e23}) and 
(\ref{e2}) following the common procedure \cite{23} by 
substitution of $p = p_0 + p'$
 and $\rho = \rho _0 + \rho '$, where $p_0$ and $\rho _0$ 
are the stationary (unperturbed) pressure and density of the 
dust particles and $p'$
 and $\rho '$
 are their perturbations, respectively. With due regard for 
the fact that $p' = c^2 \rho '$, we obtain
\begin{equation}
\frac{1}{{c^2 }}\frac{{\partial p'}}{{\partial t}} + \rho _0 
\nabla \cdot {\bf{v}} = 0 \label{e24}
\end{equation}
and
\begin{equation}
\frac{{\partial {\bf{v}}}}{{\partial t}} + \nu {\bf{v}} = - 
\frac{{\nabla p'}}{{\rho _0 }}. \label{e25}
\end{equation}
We represent the velocity in the form ${\bf{v}} = \nabla 
\psi$ to derive from (\ref{e25})
\begin{equation}
p' = - \rho _0 \left( {\frac{{\partial \psi }}{{\partial t}} + 
\nu \psi } \right). \label{e26}
\end{equation}
Substitution of (\ref{e26}) into (\ref{e24}) yields the wave 
equation
\begin{equation}
\frac{{\partial ^2 \psi }}{{\partial t^2 }} + \nu 
\frac{{\partial \psi }}{{\partial t}} = c^2 \Delta \psi , 
\label{e27}
\end{equation}
where $c$
 is the sound velocity. If the sought solution has the form 
$\psi = Ae^{i(\omega t - {\bf{k}} \cdot {\bf{r}})} $, where 
${\bf{k}}$
 is the wave vector, then we arrive at the dispersion relation 
$c^2 k^2 = \omega ^2 - i\omega \nu$ or
\begin{equation}
\frac{{ck}}{\omega } = \frac{{\sqrt {\sqrt {1 + \tilde \nu ^2 
} + 1} }}{{\sqrt 2 }} - i\frac{{\sqrt {\sqrt {1 + \tilde \nu ^2 
} - 1} }}{{\sqrt 2 }}, \label{e28}
\end{equation}
where $\tilde \nu = \nu /\omega $. If $\tilde \nu \ll 1$
 then
\begin{equation}
\frac{{ck}}{\omega } \simeq 1 + \frac{{\tilde \nu ^2 }}{8} 
- i\frac{{\tilde \nu }}{2}. \label{e29}
\end{equation}
It is seen from (\ref{e29}) that the sound velocity is not 
much different from $c$, and the damping length $c\nu /2$
 is defined by a half of the damping frequency $\nu $.

One can calculate $c$
 by differentiation of (\ref{e3}) with respect to $\rho = 
Mn_d $,
\begin{equation}
\frac{{dp^* }}{{dn^* }} = 2\Phi n^{*4/3} \left( 
{\frac{{dn^* }}{{d\Phi }}} \right)^{ - 1} + 
\frac{4}{3}\Phi ^2 n^{*1/3} . \label{e30}
\end{equation}
We differentiate $n^*$ with due regard for (\ref{e21}),
\begin{equation}
\frac{{dn^* }}{{d\Phi }} = - 8\theta n^{*2/3} e^\Phi (1 + 
\Phi ), \label{e31}
\end{equation}
to arrive at
\begin{equation}
\begin{array}{*{20}c}
  {c = \frac{{aT_e c^* }}{{e\sqrt {6M\lambda } }},\quad 
\quad \quad \quad \quad \quad \quad \quad \quad \quad 
\quad \quad \;\;} \\
  {c^{*2} = \frac{4}{3}\Phi ^2 n^{*1/3} \left[ {1 - 
\frac{3}{2}\frac{{n^{*1/3} }}{{(\Phi + 1)\left( {8 - 
3n^{*1/3} } \right)}}} \right],} \\
\end{array} \label{e32}
\end{equation}
where $n^{*1/3} = (8/3)(1 - \theta \Phi e^\Phi )$. Note that 
the sound velocity (along with the particle compressibility) 
is a continuous function at the second singular point.

Using (\ref{e12}) and (\ref{e32}) one can easily show that 
$c^2 (\Phi _c ) = 0$, $c^2 > 0$
 at $\Phi > \Phi _c $, and $c^2 < 0$
 at $\Phi < \Phi _c $. Since $n^* (\Phi )$
 is a decreasing function, in the latter case $n > n_c $. We 
can conclude that the corresponding branch of solutions of 
Eq.~(\ref{e11}) relates to an {\it unstable\/} state of the 
dust cloud with a negative compressibility. Such a state 
would tend to collapse and to quench the discharge. Hence, 
we confine ourselves to the treatment of a positive 
compressibility branch with $\Phi > \Phi _c$ and $n < n_c 
$.
\begin{figure}
\centering \unitlength=0.24pt
\begin{picture}(800,800)
\put(-115,-20){\includegraphics[width=9.4cm]{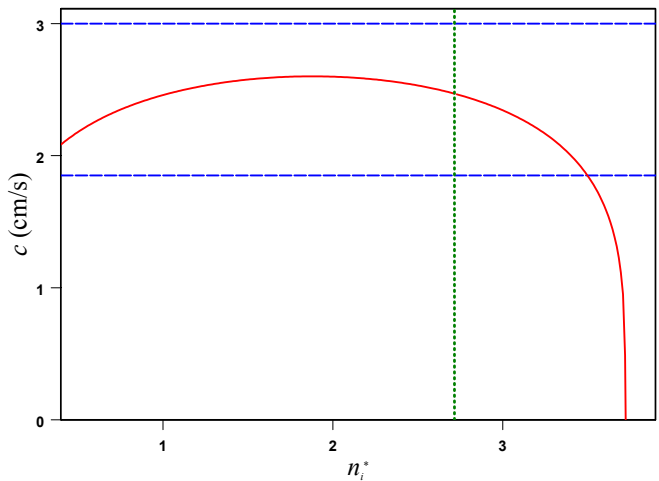}}
\end{picture}
\caption{\label{f3} (Color online) Dependence of the sound 
velocity on the dimensionless ion number density (solid 
line), $\theta = {\mbox{0}}{\mbox{.0431}}$. Dashed lines 
indicate the boundaries of a band, in which the sound 
velocity measured at different locations inside the dust 
cloud are scattered \cite{19}, and dotted line points to the 
location of the second singular point.}
\end{figure}

According to (\ref{e32}) the sound velocity is a function of 
the spatial coordinate. This dependence can be qualitatively 
illustrated if we consider the dependence $c(n_i )$
 determined by (\ref{e11}) and (\ref{e32}) (Fig.~\ref{f3}). 
For Fig.~\ref{f3}, the experimental conditions \cite{19} 
were selected as typical ones. It is seen that the sound 
velocity is almost independent of the ion number density, 
i.e., of the coordinate in the volume of the dust cloud. This 
is accounted for by the fact that the particle pressure 
(\ref{e3}) increases but the particle potential decreases with 
the increase of the dust number density. The increase in 
$n^*$ is almost compensated by the decrease in $\Phi $. 
Thus, $c(n_i )$
 has a very wide maximum with the maximum point 
situated approximately in the center of a cloud 
(Fig.~\ref{f3}). The variation of $c(n_i )$
 is so small that the entire curve lies within the band, in 
which the sound velocity measured at different locations 
inside the dust cloud is scattered \cite{19}. This accounts 
for the fact that the dependence of sound velocity on the 
coordinate was not resolved in experiments. It is also seen 
in Fig.~\ref{f3} that in the close vicinity of the inner 
boundary of the cloud, which we associate with the first 
singular point, $c$
 vanishes very sharply. Apparently, behavior of $c$
 in this region cannot be resolved experimentally as well.

Obviously, the average sound velocity is close to its value 
at the second singular point $c_s $. Using (\ref{e18}) and 
(\ref{e22}) we obtain from (\ref{e32})
\begin{equation}
\begin{array}{*{20}c}
  {c_s = \frac{{Zec_s^* }}{{\sqrt {6M\lambda } }} = 
\frac{{T_e c_s^* }}{{e\sqrt {8\pi \rho _d a\lambda } 
}},\;\;} \\
  {c_s^* = \frac{8}{3}\frac{{\Phi _s (\Phi _s + 1)}}{{\sqrt 
{(3\Phi _s + 4)(\Phi _s + 2)} }}.} \\
\end{array} \label{e33}
\end{equation}
It is seen in Fig.~\ref{f2} that $c_s^*$ is a decreasing 
function of the temperature parameter $\theta $. Since $\Phi 
_s \to \infty$ at $\theta \to 0$
 (Sec.~\ref{s3}), $c_s \to \infty$ in this limit; at $\theta \to 
\infty $, we have $\Phi _s \to 0$
 and $c_s \to 0$. As it follows from (\ref{e33}), $c_s$ is 
weakly dependent on the particle radius $a$.

\section{\label{s5}ANALYSIS OF EXPERIEMNTAL 
DATA}

\begin{figure}
\centering \unitlength=0.24pt
\begin{picture}(800,800)
\put(-115,-20){\includegraphics[width=9.4cm]{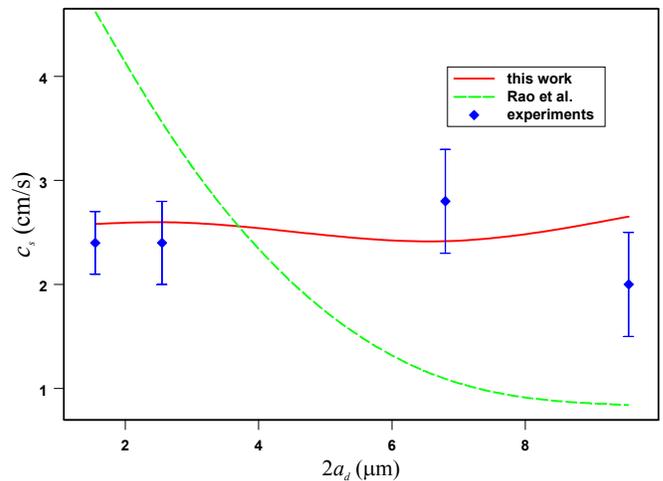}}
\end{picture}
\caption{\label{f4} (Color online) Sound velocity in the 
dust clouds formed by the particles of different diameter in 
argon RF discharge. Solid line indicates the calculation 
using (\ref{e33}); dashed line, using the formula in 
Ref.~\cite{48} (\ref{e34}). Dots represent experimental 
data for different particle diameter (from left to right): 
$2a_d = 1.55$
 \cite{019}, $2.55$
 \cite{19}, $6.8$
 \cite{11,12}, and $9.55\;\mu {\mbox{m}}$
 \cite{11,14}.}
\end{figure}

In Fig.~\ref{f4}, we compare the calculation using formula 
(\ref{e33}) with the available results of experiments, where 
the sound velocity was determined. In these experiments 
performed under microgravity conditions both in parabolic 
flights and onboard the ISS, a 3D dust cloud was formed in 
argon RF discharge. One can confirm a satisfactory 
agreement between proposed theory and experiment in a 
wide range of the particle diameter. Note a very weak 
dependence of $c_s$ on this parameter. According to 
(\ref{e33}), this follows not only from the dependence $c_s 
\sim 1/\sqrt a$ but also from $c_s \sim 1/\sqrt \lambda$ and 
from the fact that the experimental pressure is higher for 
lager particles. A good correlation between (\ref{e33}) and 
experiment is illustrated by Fig.~\ref{f3}, where the range 
of $n_i^*$ variation corresponds to the experiment 
\cite{19} (for the spatial distributions of $n_i$ and $n_d$ in 
the dust cloud; see Ref.~\cite{22}). The variation of sound 
velocity is noticeably smaller than the experimental data 
scatter.

Figure~\ref{f4} also shows the results of calculation using 
the formula $c = \sqrt {ZT_i H/M(H + 1)}$ \cite{48}. For 
the correctness of comparison, we estimate $c$
 at the second singular point. Thereby, we include the effect 
of particle charge decrease as compared to the charge of an 
isolated particle, which is sometimes called ``the charge 
cannibalism'' \cite{58,59}. Using (\ref{e023}) we obtain 
for this point
\begin{equation}
c_s = \frac{1}{{ae}}\sqrt {\frac{{3T_e T_i }}{{2\pi \rho 
_d }}\frac{{\Phi _s^2 }}{{3\Phi _s + 4}}} . \label{e34}
\end{equation}
Note that (\ref{e34}) is entirely different from (\ref{e33}). 
Figure~\ref{f4} shows that formula (\ref{e34}) 
demonstrates a trend incompatible with that of experimental 
data. Estimates show a significant variation of the sound 
velocity \cite{48} as a function of the position in the dust 
cloud. Under conditions of Ref.~\cite{19}, $c$
 would change by a factor of $1.6$, which could be 
detected in this experiment. Thus, one can confirm a better 
overall applicability of formula (\ref{e32}).

Although the linear theory of DAWs is not applicable to the 
self-excited nonlinear dust-density waves, it is of interest to 
consider their phase velocities determined in experiments 
\cite{61,62}. In the experiment \cite{61} performed in 
argon with particles of diameter $9.55\;\mu {\mbox{m}}$
 under microgravity conditions, the phase velocity varied in 
the range from $1$
 to $3\;{\mbox{cm/s}}$. In the ground-based experiment 
\cite{62}, the particles of the diameter $0.97\;\mu 
{\mbox{m}}$
 in argon were used, and the average phase velocity was 
found to be $7.5\;{\mbox{cm/s}}$. In both studies, the 
phase velocity showed considerable spatial variations. 
These results are best fitted by Eq.~(\ref{e34}) (cf.\ dashed 
line in Fig.~\ref{f4}). Apparently, a qualitative difference 
from the results concerning the sound velocity can be 
explained by the fact that the phase velocity of dust-density 
waves is defined by the ion density modulation, which is 
explicitly taken into account in (\ref{e34}) but does not 
contribute to the sound velocity in our case.
\begin{table*}
\caption{\label{t1} Parameters of a dust cloud in the PK-4 
experiments with neon complex plasma for the different 
diameters $2a$
 of monodisperse particles and their number density $n_d$ 
(experimental data were borrowed from Ref.~\cite{37}). 
The ``dust invariant'' $\kappa = r_d^2 /aT_e$ and the 
sound velocity $c_s$ (\ref{e33}) at different neon pressures 
were calculated at $T_e = 7\;{\mbox{eV}}$
 for the silica particles of $2a = 1.2\;\mu {\mbox{m}}$
 and the melamine formaldehyde particles of larger 
diameters.}
\begin{ruledtabular}
\begin{tabular}{ccccc}
$2a,\;\mu {\mbox{m}}$
 & $n_d ,\;10^4 \,{\mbox{cm}}^{ - 3}$ & $\kappa 
,\;{\mbox{cm/eV}}$
 & $c_s ,\;{\mbox{cm/s}}$
 ($p_n = 15\;{\mbox{Pa}}$) & $c_s ,\;{\mbox{cm/s}}$
 ($p_n = 30\;{\mbox{Pa}}$) \\ \hline
1.2 & 20 & 0.268 & 4.14 & 5.85 \\
6.8 & 4 & 0.138 & 1.95 & 2.76 \\
11 & 1.3 & 0.181 & 1.53 & 2.17 \\
\end{tabular}
\end{ruledtabular}
\end{table*}

In the ongoing PK-4 experiments, dust clouds are 
predominantly formed in the dc (or combined dc+RF) 
discharge in neon. First, it is necessary to check if IEOS 
used in this work is applicable for such systems. The results 
of calculation of the ``dust invariant'' $\kappa = r_d^2 
/aT_e $, which was introduced in Ref.~\cite{22} for the 
argon RF discharge, for conditions of the PK-4 experiment 
with neon is presented in Table~\ref{t1}. It is seen that 
$\kappa$ is not much different for neon and argon ($\kappa 
= 0.209\;{\mbox{cm/eV}}$
 \cite{22}). This makes it possible to apply formula 
(\ref{e33}) for the prediction of typical sound velocities in 
the PK-4 experiments. It is worth mentioning that for $2a = 
1.2\;\mu {\mbox{m}}$, $\kappa$ is twice as high as for 
$6.8\;\mu {\mbox{m}}$
 (Table~\ref{t1}). This may indicate a poor applicability of 
the approximation of similar complex plasmas for the 
smallest particles ($2a < 2\;\mu {\mbox{m}}$). The reason 
why the theory may be flawed in this range of diameters 
may lie in the fact that for the smallest particles, the Debye 
length and the momentum transfer cross section for an 
isolated particle are no longer greater than $r_d $, so that 
the Coulomb potentials of neighboring particles do not 
overlap \cite{22}. Therefore, the momentum transfer cross 
section of the ion scattering on particles may be different 
from that used in this work. Another one reason may be the 
higher electron temperature as compared to argon at low 
neon pressure. Under such conditions, the ion drag force 
can be proportional to the square of the electric field, which 
is not taken into account in used formula for the ion drag 
force (\ref{e5}).

One can also suggest the mechanisms of particle charging 
other than OML approximation, which can effectively 
change $\theta $, such as the ion-neutral collisions 
\cite{60,43}. However, as was demonstrated in 
Refs.~\cite{60,43}, the effect of collisions, which reduces 
the particle charge, is negligibly small at the pressures $p_n 
< 30\;{\mbox{Pa}}$
 and the Havnes number greater than unity. It is noteworthy 
that the typical particle charge $Z = - aT_e \Phi _c /e^2$ 
calculated using formula (\ref{e13}) for $T_e = 
7\;{\mbox{eV}}$
 and $a = 1.3\;\mu {\mbox{m}}$
 amounts to $\left| Z \right| \approx 3500$, which is almost 
half the charge of an isolated particle. This estimate is close 
to the charge determined in experiment \cite{60} performed 
at the same electron temperature and particle radius. Note 
that a particle flow rather than a stationary dust cloud was 
realized in Ref.~\cite{60}, for which (\ref{e13}) is not 
directly applicable. We also note that DAWs are unlikely to 
be resolved at $p_n > 30\;{\mbox{Pa}}$
 due to the high damping rate \cite{35}, so that inclusion of 
the effect of ion-neutral collisions would not change the 
numerical results significantly.

Thus, in PK-4 experiments, one can expect the sound 
velocity ranging from $2$
 to $3\;{\mbox{cm/s}}$
 with a weak trend to the increase with the increase in 
pressure and the decrease in particle diameter, while greater 
velocities seem to be overestimated in Table~\ref{t1}. It is 
worth mentioning that for neon, formula (\ref{e34}) would 
lead to the sound velocities ranging from $0.89$
 to $7.33\;{\mbox{cm/s}}$.

Unfortunately, no direct measurement of the sound velocity 
was performed in PK-4 experiments. However, the 
experimental data on dust acoustic shock waves are best 
fitted at a sound velocity of $2.5\;{\mbox{cm/s}}$
 \cite{50}. For melamine formaldehyde particles of 
diameter $2a = 3.4\;\mu {\mbox{m}}$
 used in this experiment, $p_n = 15\;{\mbox{Pa}}$, and 
$T_e = 7\;{\mbox{eV}}$, formula (\ref{e33}) yields a 
close value $c_s = 2.76\;{\mbox{cm/s}}$.

The sound velocity was measured for the RF discharge in 
neon in the PK-3 Plus experiment \cite{35}. For the silica 
particles of the diameter $2a = 1.55\;\mu {\mbox{m}}$, the 
neon pressure $p_n = 15\;{\mbox{Pa}}$, and the electron 
temperature $T_e = 7\;{\mbox{eV}}$
 ($\theta = {\mbox{0}}{\mbox{.0858}}$), the sound 
velocity reported in Ref.~\cite{35} is $c_s = 
0.96\;{\mbox{cm/s}}$. For this set of parameters, formula 
(\ref{e33}) yields $c_s = 3.6\;{\mbox{cm/s}}$
 and (\ref{e34}), $c_s = 5.7\;{\mbox{cm/s}}$. Despite that 
the result of (\ref{e33}) is significantly closer to the 
experiment, it is still rather inappropriate. To find the 
reason of such discrepancy, we will analyze the basic limits 
of validity of IEOS formulated in Ref.~\cite{22}. The 
Coulomb momentum transfer cross section for an isolated 
particle $ \sim (aT_e /T_i \Phi )^2$ must be much greater 
than the cross section of the particle Wigner--Seitz cell $ 
\sim r_d^2 $. At the second singular point, we have from 
(\ref{e22}) $r_d \approx 0.63\lambda $, and this condition 
can be written as
\begin{equation}
2.5\left( {\frac{{a\Phi _s T_e }}{{\lambda T_i }}} 
\right)^2 \gg 1. \label{e35}
\end{equation}
The second condition requires that the maximum impact 
parameter of ion scattering on an isolated particle with due 
regard for the Debye screening $(2\Phi a\lambda _D T_e 
/T_i )^{1/2} $, where $\lambda _D = (T_i /4\pi n_i e^2 
)^{1/2}$ is the ion Debye screening length, is much larger 
than $r_d $. At the second singular point, we use 
(\ref{e20}) and (\ref{e22}) to write this condition as
\begin{equation}
0.33\Phi _s^{1/2} \frac{{3\Phi _s + 4}}{{\Phi _s + 
1}}\left( {\frac{{aT_e }}{{\lambda T_i }}} \right)^{1/2} 
\gg 1. \label{e36}
\end{equation}
As is seen from (\ref{e35}) and (\ref{e36}), the limit of 
validity of used theory is reached at sufficiently small 
particle diameters. Under the experimental conditions 
\cite{35}, the l.h.s.\ of (\ref{e35}) and (\ref{e36}) are close 
to unity, and therefore, the theory may be flawed in this 
region.

In addition to the reasons of the inapplicability of 
(\ref{e33}) for the smallest particles discussed above, one 
can assume that in the presence of the particles in RF 
discharge, the electron temperature can be lower than that 
in the absence of the particles, i.e., less than 
$7\;{\mbox{eV}}$. However, for the experiment \cite{35}, 
there is a good agreement between the damping rates of 
DAWs first measured for a 3D dust cloud and the 
theoretical result $\nu /2$
 [see (\ref{e29})]. In fact, for $p_n = 20\;{\mbox{Pa}}$, 
the damping rates are $43$
 and $46\;{\mbox{s}}^{ - 1}$ from the theory and 
experiment, respectively, and for $p_n = 
15\;{\mbox{Pa}}$, $33$
 and $32\;{\mbox{s}}^{ - 1} $.

\section{\label{s6} CONCLUSION}

To summarize, we have calculated the sound velocity 
corresponding to DAWs in a nonequilibrium stationary 3D 
dust cloud formed in the low pressure ac-dc discharge 
under microgravity conditions. The dust cloud is assumed 
to conform to the model of similar complex plasmas, which 
treats a stationary state of the dust cloud as a balance 
between the electric force from an external electric field of 
the discharge and the ion drag force. The ionization state of 
such a system is fully defined by the one-parametric IEOS, 
which can relate each pair of four dimensionless variables. 
To find characteristic regions of behavior of the sound 
velocity, we determine two singular points of this IEOS 
related to the maximum of the ion and electron number 
densities, respectively. We use the fluid approach (the Euler 
and continuity equations) to account for the dynamics of the 
dust plasma component. We have demonstrated that in the 
presence of the external electric field, the field perturbation 
associated with the sound wave is dominated by the 
gradient of the dust pressure, which emerges due to the 
particle correlations in a strongly coupled system. Thus, the 
sound velocity is fully determined by the compressibility of 
the dust cloud. We have shown that only one branch of two 
IEOS solutions that implies the high particle potentials and 
low number densities can be realized. We have included the 
neutral drag term in the Euler equation to derive a 
dispersion relation that makes it possible to estimate the 
damping rate of a sound wave. Based on this equation, we 
calculated the dependence of the sound velocity on the ion 
number density, which mimics a real spatial distribution of 
this quantity in the dust cloud. The sound velocity was 
found to be almost independent of the coordinate and to 
assume its typical value at the second singular point.

Comparison with available experimental data reveals a 
good correspondence with the measurements of sound 
velocity. The obtained formula for this quantity can account 
for all experimentally observed regularities, such as 
independence of the coordinate, of the particle radius, and 
of the gas pressure. The damping rates of a dust cloud in 
neon are also in a good agreement with experiment.

A comparison with the well-known result of Ref.~\cite{48} 
performed at the same point of the plasma ionization state 
diagram, thus including the effect of particle charge 
lowering, shows a trend entirely different from 
experimentally observed one. For small particle diameters, 
the calculated sound velocity is more than twice as high as 
the experimental one and for large diameters, vice versa. 
This is not surprising because the result \cite{48} is quite 
correct for equilibrium ideal plasma in the absence of an 
external electric field. In contrast, our system is strongly 
nonequilibrium and strongly coupled.

In regard to the PK-4 experiment, the analyses of up-to-date 
data demonstrates that the dust cloud in neon can be 
quantified by the ``dust invariant'' of our model, which 
have approximately the same value for different dust clouds 
in neon. Its value is close to that typical for PK-3 Plus 
experiments. This allows one to make some predictive 
estimations of the sound velocity, which can be helpful in 
the analysis of dust acoustic shock wave propagation. A 
best fit sound velocity that follows from such analysis 
\cite{50} almost coincides with that calculated in this work.

A single experimental datum, which proved to be 
incompatible with the proposed theory, is a low sound 
velocity in the PK-3 Plus experiment with neon \cite{35}, 
which is more than twice as low as the theoretical 
estimation. We show that our theory is valid for sufficiently 
large particles but the dust particles used in the experiment 
\cite{35} are too small for the theory. A correct approach to 
the theory of a cloud of small particles requires treatment of 
the Yukawa rather than the Coulomb system, although the 
corresponding model may not have the property of 
similarity. Another problem of interest is the sound 
propagation in a cloud with different particle sizes. These 
problems will be addressed in the future.

\begin{acknowledgments}
This research is supported by the Russian Science 
Foundation Grant No.~14-50-00124.
\end{acknowledgments}

\providecommand{\noopsort}[1]{}\providecommand{\singleletter}[1]{#1}%

\end{document}